\documentclass[conference]{IEEEtran}
\IEEEoverridecommandlockouts
\usepackage{cite}
\usepackage{amsmath,amssymb,amsfonts}
\usepackage{algorithmic}
\usepackage{graphicx}
\usepackage{multirow}
\usepackage{textcomp}
\usepackage{xcolor}
\usepackage{booktabs}
\def\BibTeX{{\rm B\kern-.05em{\sc i\kern-.025em b}\kern-.08em
    T\kern-.1667em\lower.7ex\hbox{E}\kern-.125emX}}

\usepackage{comment}

\begin{document}

\newcommand{\new}[1]{\textcolor{red}{#1}}
\newcommand{\mozhgan}[1]{\textcolor{black}{#1}}

\title{MedMambaLite: Hardware-Aware Mamba for Medical Image Classification}

\author{
\IEEEauthorblockN{Romina Aalishah, Mozhgan Navardi, and Tinoosh Mohsenin}
\IEEEauthorblockA{
Department of Electrical and Computer Engineering\\
Johns Hopkins University, Baltimore, MD, USA\\
\{raalish1, mnavard1, tinoosh\}@jhu.edu
}
}

\maketitle

\begin{abstract}
AI-powered medical devices have driven the need for real-time, on-device inference in healthcare domains such as biomedical image classification. 
Deployment of deep learning models at the edge is now used for applications such as anomaly detection and classification in medical images. However, achieving this level of performance on edge devices remains challenging due to limitations in model size and computational capacity. To address this, we present MedMambaLite, a 
hardware-aware Mamba-based model 
optimized through knowledge distillation for medical image classification. We start with a powerful MedMamba model, which integrates a Mamba structure for efficient feature extraction in medical imaging. We make the model lighter and faster in training and inference by reducing the redundancies in the architecture and integrating modifications. We then distill its knowledge into a smaller student model by reducing the embedding dimensions. The 
optimized model achieves 94.5\% overall accuracy on 10 MedMNIST datasets. It also reduces parameters 22.8$\times$ compared to MedMamba. Deployment of MedMambaLite on an NVIDIA Jetson Orin Nano achieves 
35.6 GOPS/J energy per inference. This outperforms MedMamba by 
63\% improvement in energy per inference, demonstrating its suitability for edge medical applications. 
\end{abstract}


\section{Introduction}
Artificial Intelligence~(AI) 
in medical applications have gained attention recently for their ability to provide fast, reliable results~\cite{medcl, biasmedqa, ray2024vit, walczak2025eden}. 
These models learn the detailed patterns and structures of the anomalies and radiology images and use them to aid in early diagnosis and clinical decision-making. Image classification is one such task in computer vision and medical applications. The goal is to provide an accurate prediction to improve the speed of diagnosis.
However, deploying these models on edge devices might be challenging because of their complexity and high intensiveness by the need to capture
fine-grained features in medical images. Since edge devices are constrained in several aspects such as computing resources and power, the importance of 
number of parameters, floating-point operations (FLOPs) and reliability comes into matter~\cite{aalishah2025mambalitesr, coughnetv2, tinymnetv3, yolo, mozhgan-2024-ESL-metatinyml, mozhgan-2025-AAAI-SSS-genai, micro2023-mozhgan, pourmehrani2024fat}. 

Therefore, many Deep Learning~(DL)-based solutions have been developed to improve the accuracy and efficiency of disease detection and treatment~\cite{yue2024medmamba}. 
Early classification approaches relied on Convolutional Neural Networks~(CNNs)~\cite{unet, cnnmedimg, anthimopoulos2016ild}. While small, these CNN-based methods demonstrated relatively low accuracy. Therefore, efforts have been made to improve the performance by incorporating residual learning such as ResNet, noise-robust strategies, and suppress irrelevant features~\cite{he2016, wang2019aleatoric, roy2019recalibrating, iscas2023-mozhgan}. Recently, research has focused on developing stronger models with novel architectures and hybrid models~\cite{medvit, yue2024medmamba, ho2025litemamba}. 
With the introduction of Mamba~\cite{mamba}, a revolution happened in optimizing feature extraction. This architecture was quickly adopted across various vision tasks, including image classification and object detection~\cite{visionmamba, mambavision}. Having similar results to transformer-based architectures but with better processing speed, led to its adoption in medical applications, where it involved combining the structure of Mamba with transformer-based modules~\cite{yue2024medmamba, vm-unet, bansal2024mamba}, resulting in fast and accurate models. However, 
model deployment on edge devices is still challenging because of its size and power usage.

\begin{figure}
    \centering
    \includegraphics[width=0.5\textwidth]{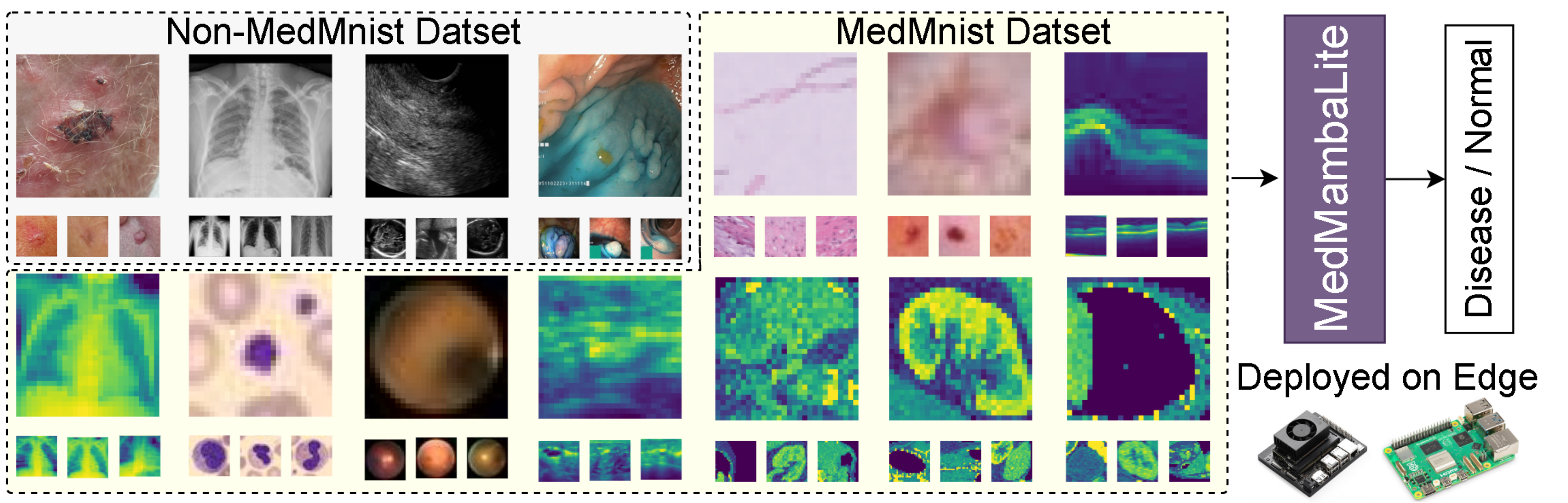}
    \vspace{-10pt}
    \caption{The overall flow of MedMambaLite. A related image to any of the 14 medical datasets is passed to the corresponding MedMambaLite model. The model outputs the class, which is either the type of disease or normal.}
    \label{fig:intro}
    \vspace{-16pt}
\end{figure}

To address these limitations, we propose MedMambaLite, a hardware-aware model designed for edge deployment in medical image analysis which is shown in Fig.~\ref{fig:intro}. Starting from the MedMamba model~\cite{yue2024medmamba}, a Mamba-based model adapted to medical imaging, we make it more efficient and faster in training and inference by reducing redundancies in the architecture and integrating modifications. Then, We 
apply knowledge distillation~(KD)~\cite{hinton2015distilling} to compress it into a smaller student model. This design addresses the memory and computational constraints often faced by transformer‐based approaches. We further validate the deployment of MedMambaLite on resource-constrained edge devices such as NVIDIA Jetson Orin Nano with 8 GB memory~\cite{nvidia_jetson_orin_nano, scalcon2024ai} and Raspberry Pi 5 with 16 GB memory~\cite{raspberrypi}. 
The experimental results show that MedMambaLite fits within a lower-level cache and achieves significant
efficiency with minimal loss in performance.
Our contributions are summarized as follows:
\begin{itemize} 
    \item Efficient reconstruction of 
    MedMamba 
    into an edge-friendly model 
    by optimizing the Mamba architecture, convolution layers, and redundancy reductions. 
    \item Development of an accurate, 
    hardware-aware student model, MedMambaLite-ST, for medical image classification through knowledge distillation.
    \item Power and latency analysis for the proposed MedMambaLite-ST on edge devices such as the Raspberry Pi 5 and NVIDIA Jetson Orin Nano with Arm Cortex processors reveals the impact of memory hierarchy and onboard data transmission overhead.
    \item Real-world validation and demonstration of improved latency and energy efficiency of proposed model compared to existing methods
    in edge medical image classification. 
\end{itemize}

\begin{figure}
    \centering
    \includegraphics[width=0.5\textwidth]{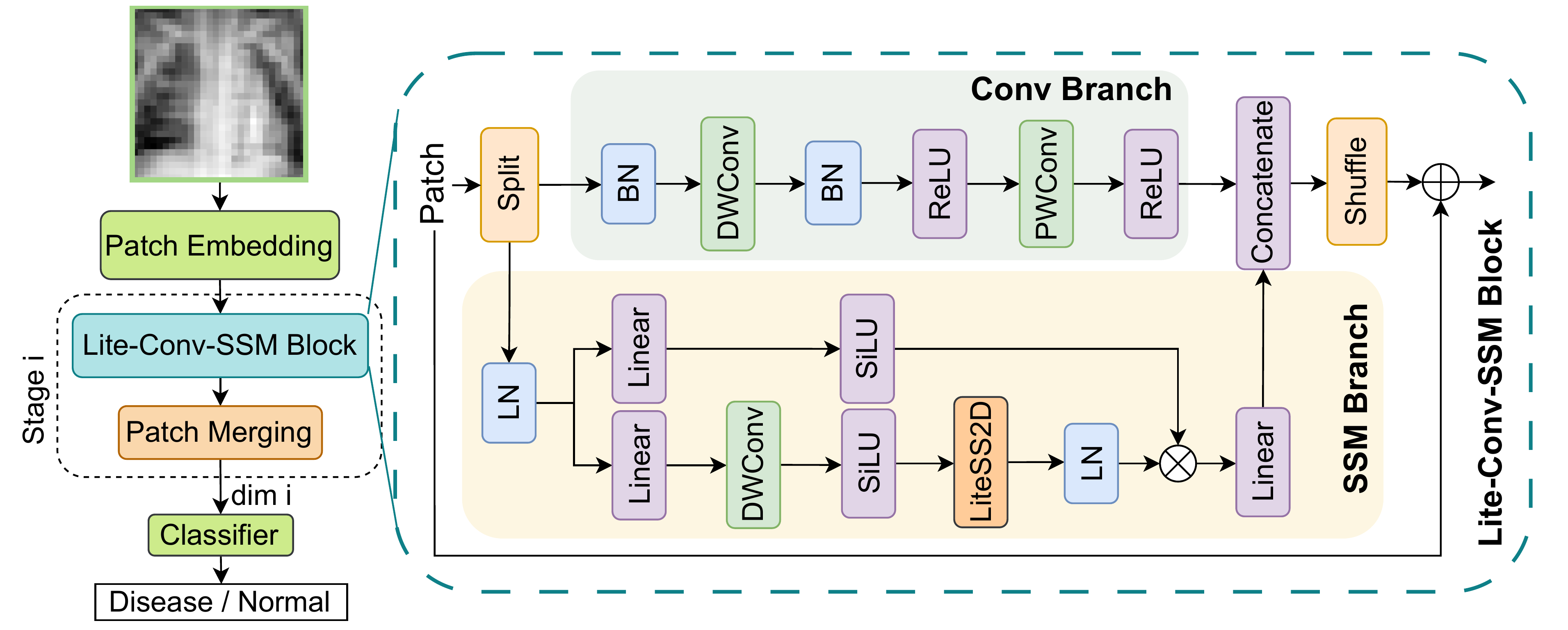}
    \caption{The proposed architecture of MedMambaLite, consisting of two branches of convolution and SSM for feature extraction. The same architecture 
    is used for teacher and student models 
    with a different set of dimensions.}
    
    \label{fig:architecture}
    \vspace{-10pt}
\end{figure}

\section{Proposed Approach}
\label{SEC:PROPOS}




\subsection{Architecture Design for Medical Image Classification}

Fig.~\ref{fig:architecture} shows the overview of the proposed architecture for medical image classification, inspired by MedMamba~\cite{yue2024medmamba}, which includes five main units as follows:

\subsubsection{Patch Embedding}

The input image is first passed through a patch embedding layer, which splits it into patches and projects them into a higher-dimensional space. 

\subsubsection{Lite-Conv-SSM-Block}
These features are then processed by a series of Lite-Conv-SSM blocks, including convolutional and State-Space Modeling~(SSM) components. Each block splits the feature map channel-wise into two halves: one is passed through a convolutional branch for capturing local features, while the other is passed through an SSM branch, utilizing a Lite 2D Selective Scan Mamba (LiteSS2D) module to capture long-range dependencies and global features. 

The convolution branch consists of depthwise and pointwise convolutions to avoid the high computations of standard ones, leading to a reduction in FLOPs.
The SSM branch runs features through a depthwise convolution, then applies a shared selective scan across four directions to efficiently capture long-range patterns.
These two branches are connected again by concatenation and channel shuffle operations, which help to make sure local and global features are not kept in only one branch. These altogether construct the Lite-Conv-SSM block. The network is organized into several stages, with each stage containing a fixed number of Lite-Conv-SSM blocks. 
This structure allows MedMambaLite to extract both fine and global features efficiently.

\subsubsection{Lite 2D-Selective-Scan}

Fig.~\ref{fig:litess2d} shows the 
architecture of the proposed Lite 2D-Selective-Scan (LiteSS2D), 
which improves the efficiency, without sacrificing performance. Therefore, instead of maintaining separate projection layers and parameters for each scan direction, LiteSS2D shares a single set of weights across all four directions, horizontal, vertical, and their reversals, leading to parameter count and computation reduction. 
The block starts by projecting the input features into a higher dimension using a linear layer, then applies separable depthwise convolutions first along rows, then along columns, with a SiLU nonlinearity in between.
The features are then passed to a four-way Selective State Space (S6) scan module, which performs direction-aware sequential modeling:

\begin{figure}
    \centering
    \centerline{\includegraphics[width=0.465\textwidth]{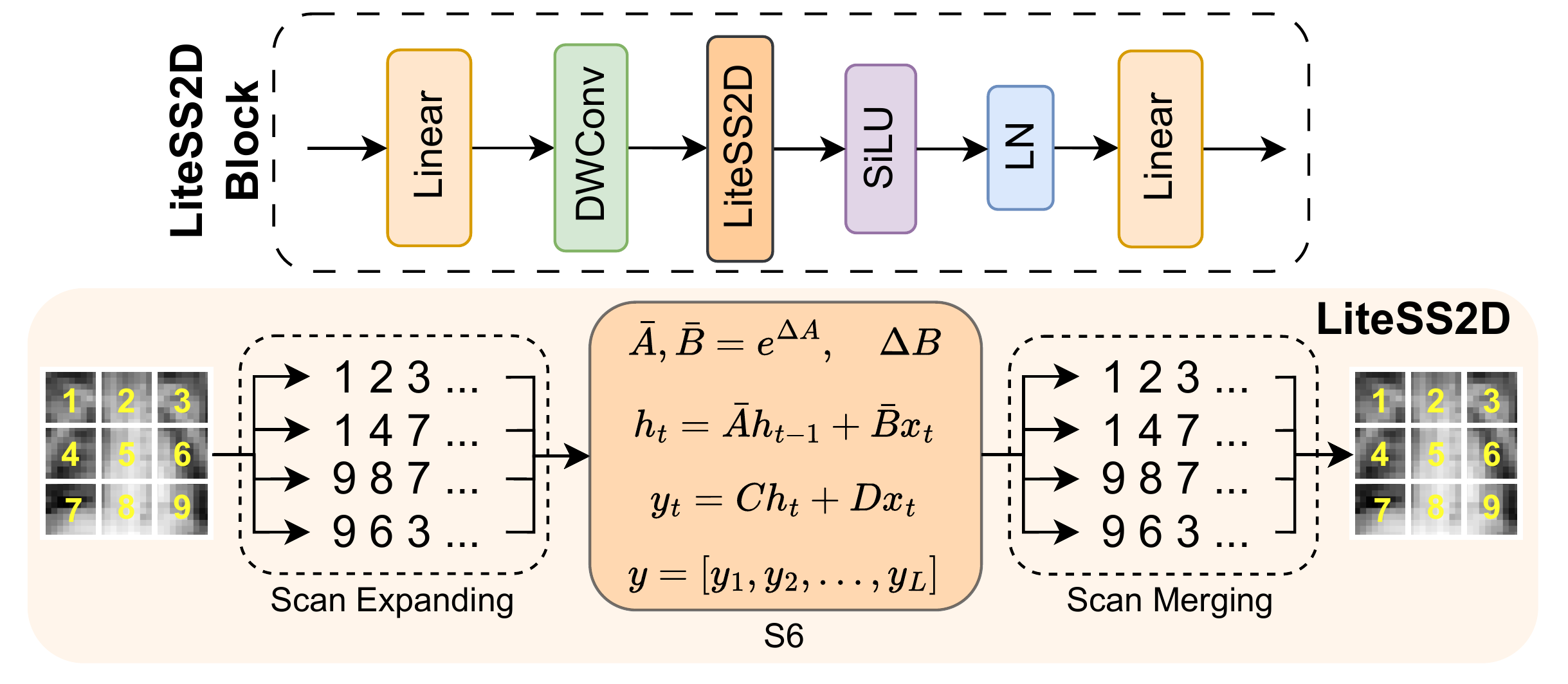}}
    \caption{The architecture of LiteSS2D, following the overall flow of the SS2D proposed by VMamba~\cite{liu2024vmamba}, but with modifications for better efficiency.}
    \label{fig:litess2d}
    \vspace{-12pt}
\end{figure}

\textbf{Scan Expanding:} the spatial tensor is flattened along four directions (horizontal, vertical, and their reversed versions) to form sequential inputs.

\textbf{S6 block:} a shared SSM core processes each sequence with the following recurrence:

\begin{equation}
\begin{aligned}
\Delta_t &= W_\Delta v_t, \quad B_t = W_B v_t, \quad C_t = W_C v_t, \\
h_t &= A_d h_{t-1} + B_d v_t, \quad A_d = \exp(A_{\log}) \\
y_t &= C h_t + D v_t.
\end{aligned}
\end{equation}

Where $v_t$ represents the input and $y_t$ is the output at time step $t$. $\Delta_t, B_t, C_t$ are intermediate values derived from the input $v_t$ using learnable weights $W_\Delta, W_B, W_C$, contributing to the dynamic behavior of the state space model. $h_t$ is the hidden state. $A_d$ is the discretized state transition matrix, derived from $A_{\log}$, which is the learnable logarithmic form of $A_d$. $D$ is a feedthrough matrix, connecting input and output directly.

\textbf{Scan Merging:} the four directional outputs are summed and reshaped back to the primary format.
The output is then normalized and projected. 

This approach provides memory-efficiency by avoiding repeated tensor reshaping and using compact representations. Therefore, compared to MedMamba~\cite{yue2024medmamba}, we introduced a few important changes to reduce its size. First, instead of using regular square convolutions, we broke them down into two smaller steps: a row-wise and a column-wise convolution. This approach keeps the same functionality but reduces the number of parameters. Next, in the state-space layers (S6), we used the same projection weights across different groups instead of having a separate set for each one. This avoids repeating the same kind of computations and saves memory. Finally, we shared two key matrices, $A_{\log}$, which is the learnable logarithmic form of the state transition matrix $A_d$,  and $D$, across all blocks in the same stage, instead of assigning a separate copy to each block. These adjustments led to a lighter version of the original MedMamba.

\begin{table}
\centering
\caption{Proposed KD details: number of layers and output dimension for each of the four stages is shown, demonstrating the similar configurations, but different dimensions.}
\renewcommand{\arraystretch}{1.2}
\setlength{\tabcolsep}{4pt} 
\begin{tabular}{|c|cc|cc|cc|cc|}
\hline
\multirow{2}{*}{\textbf{Model}} & \multicolumn{2}{c|}{\textbf{i = 1}} & \multicolumn{2}{c|}{\textbf{i = 2}} & \multicolumn{2}{c|}{\textbf{i = 3}} & \multicolumn{2}{c|}{\textbf{i = 4}} \\
\cline{2-9}
 & \#layers & dim & \#layers & dim & \#layers & dim & \#layers & dim \\
\hline
Teacher & 2 & 96 & 2 & 192 & 4 & 384 & 2 & 768 \\
\hline
Student & 2 & 32 & 2 & 64 & 4 & 128 & 2 & 256 \\
\hline
\end{tabular}
\vspace{-14pt}
\label{tab:knowledge_distillation}

\end{table}

\subsubsection{Patch Merging}
Between stages, patch merging layers reduce spatial resolution while increasing channel depth, building a hierarchical representation. 

\subsubsection{Classifier}
The classifier determines whether a disease is present in the input image or if the image is normal.

\begin{table*}[t]
  \centering
  
  \caption{Parameter Savings by Each Action: component-wise comparison between MedMamba and MedMambaLite. All sizes are computed with 32-bit floating-point weights. Overall accuracies are on MedMNIST datasets. Savings are compared to MedMamba-T~\cite{yue2024medmamba}.}
  \vspace{-8pt}
  \label{tab:component_breakdown}
  \renewcommand{\arraystretch}{1.15}
  \setlength{\tabcolsep}{3pt}
  \resizebox{0.95\textwidth}{!}{%
  \begin{tabular}{lrrrrrrrrrrrrrr}
    \toprule
    & \multicolumn{4}{c}{\textbf{MedMamba-T~\cite{yue2024medmamba}}} &
      \multicolumn{5}{c}{\textbf{MedMambaLite-TR}} &
      \multicolumn{5}{c}{\textbf{MedMambaLite-ST}} \\
    \cmidrule(lr){2-5} \cmidrule(lr){6-10} \cmidrule(lr){11-15}
    \textbf{Block} &
      Params & Size & FLOPs & Acc. (\%) &
      Params & Size & FLOPs & Acc. (\%) & Saving (\%) &
      Params & Size & FLOPs & Acc. (\%) &
      Saving (\%) \\
    \midrule
    \midrule
    \textbf{Full model}       & \textbf{14.5 M} & \textbf{55.2 MB} & \textbf{2.3 G} & \textbf{95.4} & \textbf{2.4 M} & \textbf{9.0 MB} & \textbf{475.2 M} & \textbf{94.8} & \textbf{83.6} & \textbf{635.4 K} & \textbf{2.4 MB} & \textbf{154.1 M} & \textbf{94.5} & \textbf{95.9} \\
    \midrule
    Conv branch      & 36.5 k & 139.2 kB & 11.5 M & - & 19.5 k & 74.2 kB & 5.1 M & - & 46.6 & 9.7 k & 37.1 kB & 2.6 M & - & 73.4 \\
    S6 $x_{proj}$+$dt$   & 946.9 k & 3.6 MB & 236.0 M & - & 132.0 k & 503.4 kB & 141.3 M & - & 86.1 & 53.1 k & 202.4 kB & 62.6 M & - & 94.4 \\
    $A_{\log}$ + $D$        & 248.1 k & 946.3 kB & 163.8 M & - &  41.3 k & 157.7 kB & 109.2 M & - & 83.3 & 20.7 k & 78.9 kB & 54.6 M & - & 91.7 \\
    \midrule
    \textbf{Subtotal}       &\textbf{1.2 M} &\textbf{4.7 MB} & \textbf{411.3 M} & \textbf{-} &  \textbf{192.8 k}&\textbf{735.3 kB}& \textbf{255.6 M} & \textbf{-} & \textbf{83.9} & \textbf{83.5 k} & \textbf{318.4 kB} & \textbf{119.8 M} & \textbf{-} & \textbf{93.0} \\
    \bottomrule
  \end{tabular}}
 \vspace{-8pt}
\end{table*}


\begin{table*}[]
\centering

\caption{\mozhgan{Experimental results comparing MedMamba~\cite{yue2024medmamba}, MedMambaLite-TR, and MedMambaLite-ST on three platforms—Lambda server, Jetson Orin Nano (8GB), and Raspberry Pi 5 (16GB)—demonstrate the effectiveness of the proposed approach. Specifically, MedMambaLite-ST achieves up to 63\% improvement in energy efficiency on Jetson Orin Nano and 74\% on Raspberry Pi 5, highlighting the advantage of model optimization in resource-constrained edge environments.}} 
\vspace{-6pt}
\label{tab:hw}
\label{tab:hw}
\resizebox{\textwidth}{!}{%
\begin{tabular}{|l|ccc|ccc|ccc|}
\hline
Platform                   & \multicolumn{3}{c|}{Lambda Server}                                                                                                                                                                                                                                                               & \multicolumn{3}{c|}{NVIDIA Jetson Orin Nano with 8 GB Memory}                                                                                                                                                                                                                                    & \multicolumn{3}{c|}{Raspberry Pi 5 with 16 GB Memory}                                                                                                                                                                                                                                            \\ \hline
Model                      & \multicolumn{1}{c|}{\begin{tabular}[c]{@{}c@{}}MedMamba-T~\cite{yue2024medmamba} \end{tabular}} & \multicolumn{1}{c|}{\begin{tabular}[c]{@{}c@{}}MedMambaLite-TR  \end{tabular}} & \begin{tabular}[c]{@{}c@{}}MedMambaLite-ST \end{tabular} & \multicolumn{1}{c|}{\begin{tabular}[c]{@{}c@{}}MedMamba~\cite{yue2024medmamba} \end{tabular}} & \multicolumn{1}{c|}{\begin{tabular}[c]{@{}c@{}}MedMambaLite-TR  \end{tabular}} & \begin{tabular}[c]{@{}c@{}}MedMambaLite-ST \end{tabular} & \multicolumn{1}{c|}{\begin{tabular}[c]{@{}c@{}}MedMamba~\cite{yue2024medmamba} \end{tabular}} & \multicolumn{1}{c|}{\begin{tabular}[c]{@{}c@{}}MedMambaLite-TR \end{tabular}} & \begin{tabular}[c]{@{}c@{}}MedMambaLite-ST \end{tabular} \\ \hline
Power (W)                  & \multicolumn{1}{c|}{N/A}                                                                   & \multicolumn{1}{c|}{N/A}                                                                                    & N/A                                                                                   & \multicolumn{1}{c|}{\textbf{3.2}}                                                          & \multicolumn{1}{c|}{3.3}                                                                                    & \textbf{2.7}                                                                          & \multicolumn{1}{c|}{6.7}                                                                   & \multicolumn{1}{c|}{6.1}                                                                                    & 7.7                                                                                   \\ \hline
Throughput (Inference/Sec) & \multicolumn{1}{c|}{149.3}                                                                 & \multicolumn{1}{c|}{185.2}                                                                                  & 192.3                                                                                 & \multicolumn{1}{c|}{34.2}                                                                  & \multicolumn{1}{c|}{63.7}                                                                                   & 76.7                                                                                  & \multicolumn{1}{c|}{\textbf{4.3}}                                                                   & \multicolumn{1}{c|}{9.1}                                                                                    & \textbf{18.9}                                                                                  \\ \hline
Performance (GFOPS)        & \multicolumn{1}{c|}{338.1}                                                                 & \multicolumn{1}{c|}{86.9}                                                                                   & 29.2                                                                                  & \multicolumn{1}{c|}{77.4}                                                                  & \multicolumn{1}{c|}{29.9}                                                                                   & 11.7                                                                                  & \multicolumn{1}{c|}{9.8}                                                                   & \multicolumn{1}{c|}{4.2}                                                                                    & 2.9                                                                                   \\ \hline
Energy Efficiency (GOPS/J) & \multicolumn{1}{c|}{N/A}                                                                   & \multicolumn{1}{c|}{N/A}                                                                                    & N/A                                                                                   & \multicolumn{1}{c|}{814.5}                                                       & \multicolumn{1}{c|}{571.2}                                                                                  & 327.6                                                                       & \multicolumn{1}{c|}{6.3}                                                                   & \multicolumn{1}{c|}{6.3}                                                                                    & 7.1                                                                                   \\ \hline
Energy/Infrence (mJ)       & \multicolumn{1}{c|}{N/A}                                                                   & \multicolumn{1}{c|}{N/A}                                                                                    & N/A                                                                                   & \multicolumn{1}{c|}{\textbf{95.0}}                                                                  & \multicolumn{1}{c|}{52.3}                                                                                   & \textbf{35.6 }                                                                                 & \multicolumn{1}{c|}{\textbf{1560.0}}                                                       & \multicolumn{1}{c|}{678.5}                                                                                  & \textbf{405.6}                                                                        \\ \hline
\end{tabular}}
\vspace{-12pt}
\end{table*}

\subsection{Memory-aware model optimization: MedMambaLite-ST}


\mozhgan{Memory-aware model compression reduces deep learning model size with minimal performance loss. In this work, we introduce two variants of our MedMambaLite architecture: MedMambaLite-TR (teacher) and MedMambaLite-ST (student). Through knowledge distillation, the student model learns from the teacher, resulting in a compact and memory-efficient model suitable for real-time deployment on edge devices with limited cache resources.}

\subsubsection{Memory Hierarchy Analysis of Edge Devices}
\begin{figure}
    \centering
    \includegraphics[width=.4\textwidth]{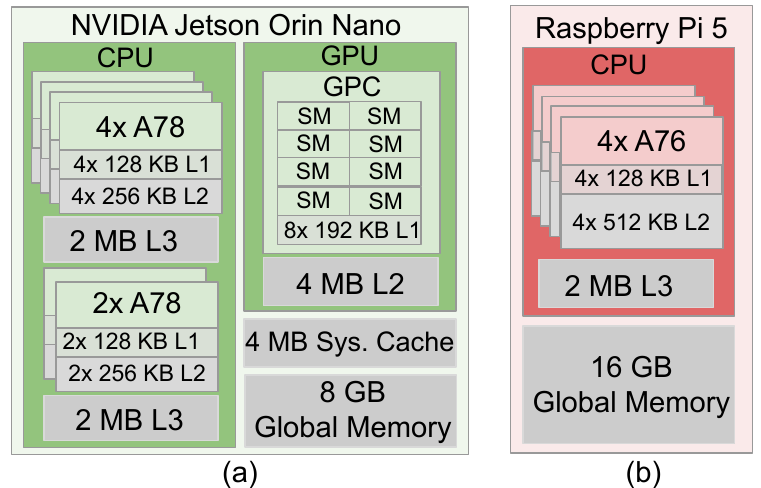}
    \vspace{-5pt}
    \caption{Hardware architecture: (a)~Jetson Orin Nano~\cite{nvidia_orin_nano_carrier_board} and (b)~Raspberry Pi 5~\cite{raspberrypi}, highlighting L1/L2/L3 cache hierarchy and available on-chip memory.}
    \label{fig:arch}
\end{figure}

\mozhgan{Fig.~\ref{fig:arch} illustrates the memory architecture and constraints of two edge devices, Jetson Orin Nano and Raspberry Pi 5, highlighting cache hierarchy, memory bandwidth, and data flow relevant to model deployment. The Orin Nano features a 6-core Cortex-A78AE CPU and an Ampere-based GPU, with 8×192 KB L1 cache, 4 MB shared L2, 4 MB system-level cache, and 8 GB global memory. Its GPU has approximately 5.5 MB of on-chip memory, enabling efficient inference. In contrast, the Raspberry Pi 5 includes a quad-core Cortex-A76 CPU with 128 KB L1, 512 KB L2 per core, 2 MB shared L3, and 16 GB global memory, totaling 4.5 MB of on-chip memory. This architectural distinction plays a critical role in determining how effectively compressed models like MedMambaLite-ST can utilize on-chip memory for low-latency, energy-efficient execution, especially under strict resource constraints.
Based on this analysis, KD is applied to compress the MedMambaLite-TR model into MedMambaLite-ST, optimizing it to fit within memory constraints while maintaining performance.}

\begin{figure}
    \centering
    \centerline{\includegraphics[width=0.4\textwidth]{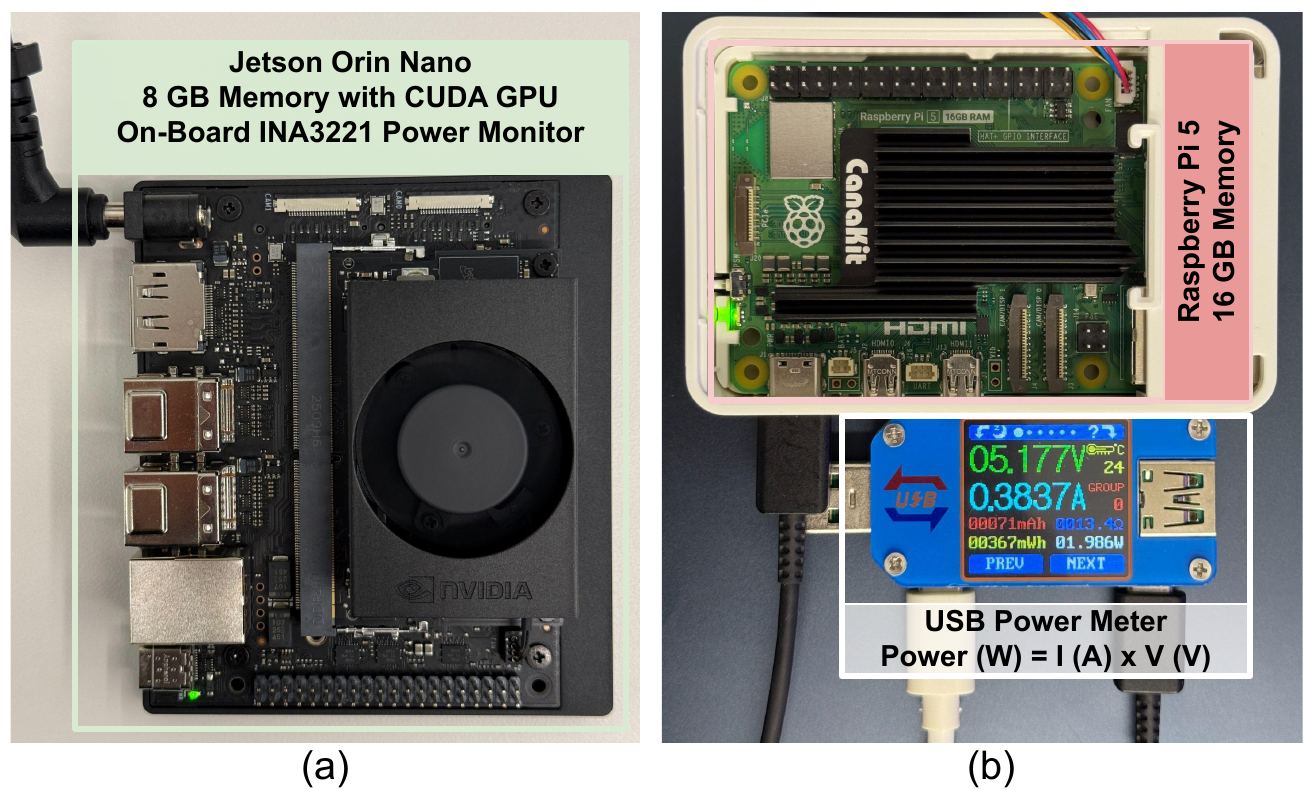}}
    \vspace{-5pt}
    \caption{Power measurement setup: (a) A USB power measurement device was used for the Raspberry Pi 5~\cite{rashid2023hac}. (b) The on-board INA3221 power monitor and tegrastats software were used for the NVIDIA Jetson Orin Nano~\cite{nvidia_power_optimization_jetson}.}
    \vspace{-14pt}
    \label{fig:hw}
\end{figure}

\subsubsection{ Knowledge Distillation for Memory-Aware Optimization}
To improve efficiency, we apply knowledge distillation~(KD) and train a student model with smaller dimensions and the 
architecture as the teacher model. This reduction in dimensions reduces the model size and number of parameters. As large teacher models can introduce computational inefficiencies and slow down the distillation process due to their high FLOPs and memory demands~\cite{hinton2015distilling,gou2021knowledge}, we adopt MedMambaLite-TR as a more efficient teacher. Table~\ref{tab:knowledge_distillation} shows the 
configuration of MedMambaLite-TR and  MedMambaLite-ST which is based on the proposed architecture for medical image classification, MedMambaLite. 
Meanwhile, KD allows the student model, MedMambaLite-ST, to mimic the performance of the teacher, MedMambaLite-TR, while being more computationally efficient. The proposed MedMambaLite-ST is optimized to retain essential features while achieving comparable accuracy. Note that both the teacher and student models receive the same input and follow the same pre-processing pipeline. \mozhgan{With the proposed MedMambaLite-ST, we achieve lower memory requirements and are able to fit the model within the L2 cache without sacrificing accuracy.}

\section{Experimental Result}
\label{SEC:EXPR}

\subsection{Experimental Setup}

\textbf{Implementation Details.}
\mozhgan{To evaluate the proposed MedMambaLite, we consider MedMamba-T, the smallest model variant presented in MedMamba~\cite{yue2024medmamba}, as the baseline. We deploy MedMambaLite-ST, MedMambaLite-TR, and the baseline on two edge devices: NVIDIA Jetson Orin Nano (8GB) and Raspberry Pi 5 (16GB), as shown in Fig.~\ref{fig:hw}. We conduct comprehensive experiments to measure latency and energy efficiency across all models.}

\textbf{Datasets.}
This study evaluates model performance using 10 datasets from MedMNIST v2~\cite{medmnist} and four 
medical imaging datasets~\cite{padufes,cpnxray,fetalplanes,kvasir}, covering diverse domains. All images are resized to $224 \times 224$, normalized to zero mean and unit variance, and adapted for grayscale or RGB input as required. 



\begin{figure}
    \centering
    \centerline{\includegraphics[width=0.5\textwidth]{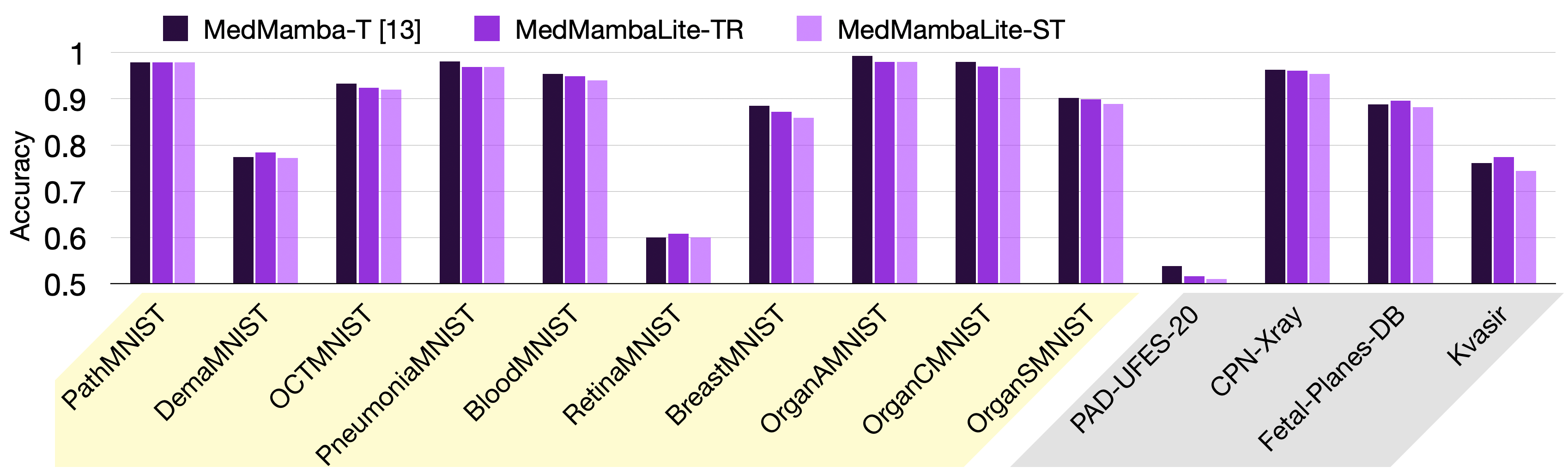}}
    \vspace{-8pt}
    \caption{\mozhgan{Overall Accuracy on selected MedMNIST (first 10) and Non-MedMNIST (last four) datasets for MedMamba variants. MedMamba-T is replicated for fair comparison. MedMambaLite-TR and MedMambaLite-ST are the proposed teacher and student models, respectively. Reported accuracies of MedMambaLite-TR and MedMambaLite-ST are on the validation set.}}
    \label{fig:per_oa}
    \vspace{-12pt}
\end{figure}

\vspace{-2.5pt}

\subsection{Results}


In knowledge distillation (KD), the teacher is actively involved throughout the training process, and using MedMamba~\cite{yue2024medmamba} results in higher resource consumption. Therefore, we introduce MedMambaLite-TR as a more efficient teacher model. Table~\ref{tab:component_breakdown} represents measurements of MedMambaLite variants after optimization  
and KD. 
Compared to the baseline MedMamba~\cite{yue2024medmamba}, the proposed teacher model, MedMamba-TR, achieves parameter reduction through several optimizations: modifying the convolutional structure cuts nearly half of the parameters; reusing projection weights in the state-space layers removed parameters 7.2×; and sharing the $A_{\log}$ and $D$ matrices across stages results in 6× reduction of parameters.
Using MedMambaLite-TR as the teacher, we further distill the model into MedMambaLite-ST. The proposed student model, MedMambaLite-ST, reduced FLOPs 14.9$\times$, the number of parameters 22.8$\times$ 
compared to the baseline.



\begin{figure}
    \centering
    \centerline{\includegraphics[width=0.4\textwidth]{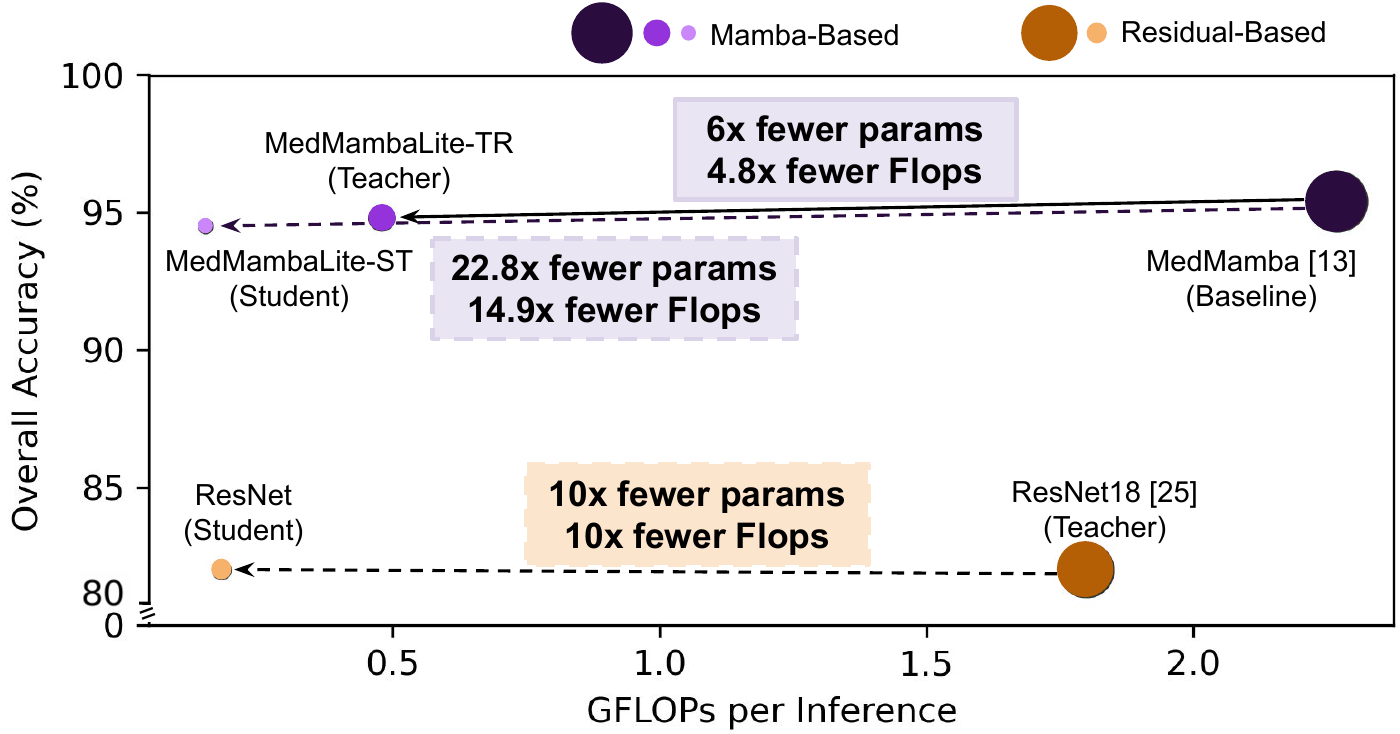}}
    \vspace{-10pt}
    \caption{\mozhgan{MedMNIST overall accuracy and FLOPs comparison between teacher and student models for Residual-based and Mamba-based architectures. 
    Smaller bubbles indicate a lower number of parameters. MedMambaLite-ST achieves up to 22.8× fewer parameters and 4.9× fewer FLOPs, demonstrating the effectiveness of the proposed model for edge devices. 
    }}
    \vspace{-12pt}
    \label{fig:oa}
\end{figure}

Fig.~\ref{fig:per_oa} depicts the accuracy of the proposed MedMambaLite compared to the baseline across 14 datasets including 10 MedMNIST datasets and four NonMedMNIST datasets. The results show that the proposed model achieves competitive accuracy relative to the baseline. 
\mozhgan{Fig.~\ref{fig:oa} shows the overall accuracy and FLOPs comparison between teacher and student models for Residual-based and Mamba-based architectures on the MedMNIST dataset. The results show that Mamba-based models achieve higher accuracy compared to Residual-based models, while maintaining competitive model size and FLOPs.
}

\mozhgan{Table~\ref{tab:hw} presents the hardware implementation results on three different platforms: Lambda Server, Jetson Orin Nano, and Raspberry Pi 5. The results show that the proposed memory-aware model, MedMambaLite-ST, fits within the lower levels of available on-chip memory on edge devices, leading to up to 63\% improvement in energy per inference.}

\section{Conclusion}
\mozhgan{In this work, a memory-aware optimization model, MedMambaLite-ST, is proposed for medical image classification. The proposed approach reduces model complexity to fit within the lower cache levels of edge devices such as the NVIDIA Jetson Orin Nano and Raspberry Pi 5, aiming to improve energy efficiency and throughput. To achieve this, memory-aware knowledge distillation is applied to the proposed MedMambaLite. Experimental results show an 22.8$\times$ reduction in number of parameters while maintaining accuracy equal to 94.5\%. Hardware implementation results demonstrate up to 63\% improvement in energy per inference.}

\bibliographystyle{IEEEtran}


\begin{thebibliography}{10}
\providecommand{\url}[1]{#1}
\csname url@samestyle\endcsname
\providecommand{\newblock}{\relax}
\providecommand{\bibinfo}[2]{#2}
\providecommand{\BIBentrySTDinterwordspacing}{\spaceskip=0pt\relax}
\providecommand{\BIBentryALTinterwordstretchfactor}{4}
\providecommand{\BIBentryALTinterwordspacing}{\spaceskip=\fontdimen2\font plus
\BIBentryALTinterwordstretchfactor\fontdimen3\font minus \fontdimen4\font\relax}
\providecommand{\BIBforeignlanguage}[2]{{%
\expandafter\ifx\csname l@#1\endcsname\relax
\typeout{** WARNING: IEEEtran.bst: No hyphenation pattern has been}%
\typeout{** loaded for the language `#1'. Using the pattern for}%
\typeout{** the default language instead.}%
\else
\language=\csname l@#1\endcsname
\fi
#2}}
\providecommand{\BIBdecl}{\relax}
\BIBdecl

\bibitem{medcl}
K.~Zhang and V.~M. Patel, ``Medcl: Learning consistent anatomy distribution for scribble-supervised medical image segmentation,'' arXiv preprint arXiv:2503.22890, 2025.

\bibitem{biasmedqa}
S.~Schmidgall, C.~Harris, I.~Essien, D.~Olshvang, T.~Rahman, J.~W. Kim, R.~Ziaei, J.~Eshraghian, P.~Abadir, and R.~Chellappa, ``Evaluation and mitigation of cognitive biases in medical language models,'' \emph{npj Digital Medicine}, vol.~7, no.~1, p. 295, 2024.

\bibitem{ray2024vit}
S.~Ray, C.-Y. Lee, H.-C. Park, D.~W. Nauen, C.~Bettegowda, X.~Li, and R.~Chellappa, ``Leveraging pretrained vision transformers for automated cancer diagnosis in optical coherence tomography images,'' \emph{medRxiv}, 2024, medRxiv preprint 2024.09.26.24314445.

\bibitem{walczak2025eden}
\BIBentryALTinterwordspacing
M.~Walczak, R.~Aalishah, W.~Mackey, B.~Story, D.~L. Boothe~Jr., N.~Waytowich, X.~Lin, and T.~Mohsenin, ``Eden: Entorhinal driven egocentric navigation toward robotic deployment,'' \emph{arXiv preprint arXiv:2506.03046}, 2025. [Online]. Available: \url{https://arxiv.org/abs/2506.03046}
\BIBentrySTDinterwordspacing

\bibitem{aalishah2025mambalitesr}
R.~Aalishah, M.~Navardi, and T.~Mohsenin, ``Mambalitesr: Image super-resolution with low-rank mamba using knowledge distillation,'' in \emph{Proceedings of the International Symposium on Quality Electronic Design (ISQED)}, 2025.

\bibitem{coughnetv2}
H.-A. Rashid, M.~M. Sajadi, and T.~Mohsenin, ``Coughnet-v2: A scalable multimodal dnn framework for point-of-care edge devices to detect symptomatic covid-19 cough,'' in \emph{Proc. 2022 IEEE Healthcare Innovations and Point of Care Technologies (HI-POCT)}, Houston, TX, USA, Mar. 2022, pp. 44--47.

\bibitem{tinymnetv3}
H.-A. Rashid and T.~Mohsenin, ``Tinymnet-v3: Memory-aware compressed multimodal deep neural networks for sustainable edge deployment,'' arXiv preprint arXiv:2405.12353, 2024.

\bibitem{yolo}
E.~Humes, M.~Navardi, and T.~Mohsenin, ``Squeezed edge yolo: Onboard object detection on edge devices,'' in \emph{Advances in Neural Information Processing Systems (NeurIPS)}, vol.~36, 2023.

\bibitem{mozhgan-2024-ESL-metatinyml}
M.~Navardi, E.~Humes, and T.~Mohsenin, ``Metatinyml: End-to-end metareasoning framework for tinyml platforms,'' \emph{IEEE Embedded Systems Letters}, vol.~16, no.~4, pp. 393--396, 2024.

\bibitem{mozhgan-2025-AAAI-SSS-genai}
M.~Navardi, R.~Aalishah, Y.~Fu, Y.~Lin, H.~Li, Y.~Chen, and T.~Mohsenin, ``Genai at the edge: Comprehensive survey on empowering edge devices,'' in \emph{Proceedings of the AAAI Symposium Series}, vol.~5, no.~1, 2025, pp. 180--187.

\bibitem{micro2023-mozhgan}
M.~Navardi, E.~Humes, T.~Manjunath, and T.~Mohsenin, ``Metae2rl: Toward metareasoning for energy-efficient multi-goal reinforcement learning with squeezed edge yolo,'' \emph{IEEE Micro}, 2023.

\bibitem{pourmehrani2024fat}
H.~Pourmehrani, J.~Bahrami, P.~Nooralinejad, H.~Pirsiavash, and N.~Karimi, ``Fat-rabbit: Fault-aware training towards robustness againstbit-flip based attacks in deep neural networks,'' in \emph{2024 IEEE International Test Conference (ITC)}.\hskip 1em plus 0.5em minus 0.4em\relax IEEE, 2024, pp. 106--110.

\bibitem{yue2024medmamba}
Y.~Yue and Z.~Li, ``Medmamba: Vision mamba for medical image classification,'' arXiv preprint arXiv:2403.03849, 2024.

\bibitem{unet}
O.~Ronneberger, P.~Fischer, and T.~Brox, ``U-net: Convolutional networks for biomedical image segmentation,'' \emph{arXiv preprint arXiv:1505.04597}, 2015.

\bibitem{cnnmedimg}
Q.~Li, W.~Cai, X.~Wang, Y.~Zhou, D.~D. Feng, and M.~Chen, ``Medical image classification with convolutional neural network,'' in \emph{Proc. 2014 13th International Conference on Control, Automation, Robotics and Vision (ICARCV)}, Singapore, Dec. 2014, pp. 844--849.

\bibitem{anthimopoulos2016ild}
M.~Anthimopoulos, S.~Christodoulidis, L.~Ebner, A.~Christe, and S.~Mougiakakou, ``Lung pattern classification for interstitial lung diseases using a deep convolutional neural network,'' \emph{IEEE Transactions on Medical Imaging}, vol.~35, no.~5, pp. 1207--1216, May 2016.

\bibitem{he2016}
K.~He, X.~Zhang, S.~Ren, and J.~Sun, ``Deep residual learning for image recognition,'' in \emph{Proceedings of the IEEE Conference on Computer Vision and Pattern Recognition (CVPR)}, 2016.

\bibitem{wang2019aleatoric}
\BIBentryALTinterwordspacing
G.~Wang, W.~Li, M.~Aertsen, J.~Deprest, S.~Ourselin, and T.~Vercauteren, ``Aleatoric uncertainty estimation with test-time augmentation for medical image segmentation with convolutional neural networks,'' \emph{Neurocomputing}, vol. 338, pp. 34--45, 2019. [Online]. Available: \url{https://doi.org/10.1016/j.neucom.2019.01.103}
\BIBentrySTDinterwordspacing

\bibitem{roy2019recalibrating}
\BIBentryALTinterwordspacing
A.~G. Roy, N.~Navab, and C.~Wachinger, ``Recalibrating fully convolutional networks with spatial and channel 'squeeze and excitation' blocks,'' \emph{IEEE Transactions on Medical Imaging}, vol.~38, no.~2, pp. 540--549, 2019. [Online]. Available: \url{https://doi.org/10.1109/TMI.2018.2867261}
\BIBentrySTDinterwordspacing

\bibitem{iscas2023-mozhgan}
M.~Navardi and T.~Mohsenin, ``Mlae2: Metareasoning for latency-aware energy-efficient autonomous nano-drones,'' in \emph{2023 IEEE International Symposium on Circuits and Systems (ISCAS)}.\hskip 1em plus 0.5em minus 0.4em\relax IEEE, 2023, pp. 1--5.

\bibitem{medvit}
O.~N. Manzari, H.~Ahmadabadi, H.~Kashiani, S.~B. Shokouhi, and A.~Ayatollahi, ``Medvit: A robust vision transformer for generalized medical image classification,'' arXiv preprint arXiv:2302.09462, 2023.

\bibitem{ho2025litemamba}
\BIBentryALTinterwordspacing
Q.~Ho, T.~Tran, and V.~Pham, ``Litemamba-bound: A lightweight mamba-based model with boundary-aware and normalized active contour loss for skin lesion segmentation,'' \emph{Methods}, 2025. [Online]. Available: \url{https://www.sciencedirect.com/science/article/pii/S1046202325000118}
\BIBentrySTDinterwordspacing

\bibitem{mamba}
A.~Gu and T.~Dao, ``Mamba: Linear-time sequence modeling with selective state spaces,'' arXiv preprint arXiv:2312.00752, 2023.

\bibitem{visionmamba}
L.~Zhu, B.~Liao, Q.~Zhang, X.~Wang, W.~Liu, and X.~Wang, ``Vision mamba: Efficient visual representation learning with bidirectional state space model,'' in \emph{Proceedings of the International Conference on Machine Learning (ICML)}, 2024.

\bibitem{mambavision}
A.~Hatamizadeh and J.~Kautz, ``Mambavision: A hybrid mamba-transformer vision backbone,'' arXiv preprint arXiv:2407.08083, 2024.

\bibitem{vm-unet}
J.~Ruan, J.~Li, and S.~Xiang, ``Vm-unet: Vision mamba unet for medical image segmentation,'' arXiv preprint arXiv:2402.02491, 2024.

\bibitem{bansal2024mamba}
\BIBentryALTinterwordspacing
S.~Bansal, S.~A, M.~J. Prasath, S.~Madisetty, M.~Z.~U. Rehman, C.~S. Raghaw, G.~Duggal, and N.~Kumar, ``A comprehensive survey of mamba architectures for medical image analysis: Classification, segmentation, restoration and beyond,'' \emph{arXiv preprint arXiv:2410.02362}, 2024. [Online]. Available: \url{https://arxiv.org/abs/2410.02362}
\BIBentrySTDinterwordspacing

\bibitem{hinton2015distilling}
\BIBentryALTinterwordspacing
G.~Hinton, O.~Vinyals, and J.~Dean, ``Distilling the knowledge in a neural network,'' 2015, nIPS 2014 Deep Learning Workshop. [Online]. Available: \url{https://doi.org/10.48550/arXiv.1503.02531}
\BIBentrySTDinterwordspacing

\bibitem{nvidia_jetson_orin_nano}
{NVIDIA}, ``Jetson orin nano developer kit getting started — nvidia developer,'' \url{https://developer.nvidia.com/embedded/learn/get-started-jetson-orinnano-devkit}, 2025, accessed: 2025-06-06.

\bibitem{scalcon2024ai}
F.~P. Scalcon, R.~Tahal, M.~Ahrabi, Y.~Huangfu, R.~Ahmed, B.~Nahid-Mobarakeh, S.~Shirani, C.~Vidal, and A.~Emadi, ``Ai-powered video monitoring: Assessing the nvidia jetson orin devices for edge computing applications,'' in \emph{2024 IEEE Transportation Electrification Conference and Expo (ITEC)}.\hskip 1em plus 0.5em minus 0.4em\relax IEEE, 2024, pp. 1--6.

\bibitem{raspberrypi}
{Raspberry Pi}, ``Getting started with your raspberry pi,'' \url{https://www.raspberrypi.com/documentation/computers/getting-started.html}, 2025, accessed: 2025-06-06.

\bibitem{liu2024vmamba}
\BIBentryALTinterwordspacing
Y.~Liu, Y.~Tian, Y.~Zhao, H.~Yu, L.~Xie, Y.~Wang, Q.~Ye, J.~Jiao, and Y.~Liu, ``Vmamba: Visual state space model,'' \emph{arXiv preprint arXiv:2401.10166}, 2024. [Online]. Available: \url{https://arxiv.org/abs/2401.10166}
\BIBentrySTDinterwordspacing

\bibitem{nvidia_orin_nano_carrier_board}
{NVIDIA Corporation}, ``Jetson orin nano developer kit carrier board specification,'' \url{https://developer.nvidia.com/embedded/downloads}, 2024, accessed: 2025-06-06.

\bibitem{rashid2023hac}
H.-A. Rashid and T.~Mohsenin, ``Hac-pocd: Hardware-aware compressed activity monitoring and fall detector edge poc devices,'' in \emph{2023 IEEE Biomedical Circuits and Systems Conference (BioCAS)}.\hskip 1em plus 0.5em minus 0.4em\relax IEEE, 2023, pp. 1--5.

\bibitem{nvidia_power_optimization_jetson}
{NVIDIA Corporation}, ``Power optimization with nvidia jetson,'' \url{https://developer.nvidia.com/blog/power-optimization-with-nvidia-jetson/}, 2021, accessed: 2025-06-06.

\bibitem{gou2021knowledge}
J.~Gou, B.~Yu, S.~J. Maybank, and D.~Tao, ``Knowledge distillation: A survey,'' \emph{International Journal of Computer Vision}, vol. 129, pp. 1789--1819, 2021.

\bibitem{medmnist}
J.~Yang, Y.~Shi, and et~al., ``Medmnist v2: A large-scale lightweight benchmark for 2d and 3d biomedical image classification,'' \emph{Scientific Data}, 2023.

\bibitem{padufes}
A.~Pacheco and R.~Krohling, ``Pad-ufes-20: A skin lesion dataset composed of patient data and clinical images collected from smartphones,'' \emph{Data in Brief}, 2020.

\bibitem{cpnxray}
T.~Rahman \emph{et~al.}, ``Covid-19 pneumonia normal chest x-ray dataset,'' 2020, available at Kaggle.

\bibitem{fetalplanes}
C.~F. Baumgartner \emph{et~al.}, ``Automatic classification of standard scan planes in fetal ultrasound using cnns,'' \emph{Medical Image Analysis}, 2017.

\bibitem{kvasir}
K.~Pogorelov and et~al., ``Kvasir: A multi-class image dataset for gastrointestinal disease classification,'' \emph{ACM Multimedia Systems}, 2017.

\end{thebibliography}


\end{document}